\documentclass[preprint,aps,showpacs]{revtex4}
\usepackage{graphicx,epsf}
\usepackage{lscape}
\usepackage{citesort}
\usepackage[colorlinks,citecolor=blue,urlcolor=red,linkcolor=red]{hyperref}

\begin{document}

\title{{\Large{\bf  B$_1$B$_{s0}$K and B$_1$B$_{s1}$K strong couplings in 
three-point QCD sum rules}}}

\author{\small
\small  M. Ali Asgarian\footnote {e-mail: m.aliasgarian@ast.ui.ac.ir}}
\affiliation{Faculty of Physics, University of Isfahan, Isfahan 81746-73441, Iran}

\begin{abstract}
An improved calculation of the strong coupling constants of B$_1$B$_{s0}$K and B$_1$B$_{s1}$K vertices is presented in the framework of the three-point QCD sum rules. The coupling constants are calculated, when both the $B_{s0}(B_{s1})$ and $K$ states are off-shell. Considering the $SU_{f}(3)$ symmetry, the results are compared with the existing predictions. \par
Key words: Strong Coupling Constant, Meson, QCD Sum Rules, $SU_{f}(3)$, Off-shell, Quark.

\end{abstract}

\pacs{}

\maketitle
\section{Introduction}
There are various applications for the strong form factors and coupling constants associated with vertices that involve mesons in the QCD, which describe the low-energy interaction among heavy mesons and light mesons, are of great importance to understand the QCD long-distance dynamics. The coupling is a fundamental parameter of the effective Lagrangian of heavy meson chiral perturbative theory (HM$\chi$PT) \cite{Cheng123, Burdman123}, which plays an important role in studying heavy meson physics. At high-energy physics, it is imperative to know the exact functional form of the strong form factors in meson vertices to investigate meson interactions. More accurate determination of these coupling constants plays an important role in the understanding of the interactions of the final states in the hadronic decays of the heavy mesons. 
 The following coupling constants have been determined by different research groups:$D^* D^*\rho$\cite{MEBracco}, $D^* D \pi$ \cite{FSNavarra,MNielsen}, $D D
\rho$\cite{MChiapparini}, $D^* D \rho$\cite{Rodrigues3}, $D D
J/\psi$ \cite{RDMatheus},  $D^* D J/\psi$ \cite{RRdaSilva},
$D^*D_sK$, $D^*_sD K$, $D^*_0 D_s K$, $D^*_{s0} D K$ \cite{SLWang},
$D^*D^* P$, $D^*D V$, $D D V$ \cite{ZGWang}, $D^* D^* \pi$
\cite{FCarvalho}, $D_s D^* K$, $D_s^* D K$ \cite{ALozea}, $D D
\omega$ \cite{LBHolanda}, $D_s^*D_s\phi$ \cite{Guo123}  $D_s D_s V$, $D^{*}_s D^{*}_s V$ \cite{KJ,KJ2}, $D_1D^*\pi, D_1D_0\pi, D_1D_1\pi$ \cite{Janbazi1},
$D_{s1}D^*K $ and $D_{s1}D^*K_0^* $ \cite{Janbazi2}, $D_s DK^* $ and $ D_s D^{*} K^* $  \cite{Janbazi3},  $K^*K\pi$, $KK\phi$, $K^*K^*\phi$, $K^*K^* \rho$ \cite{Kazemi}, $D_s^*D^*K^*$  and $D_{s1}D_1K^*$ \cite{Janbazi4}, and $D^*D^*_sK$ and $D^*D_sK$ \cite{Janbazi5}, in the framework of three-point QCD sum rules. It is very important to know the precise functional form of the form factors in
these vertices and even to know how this form changes when one or
the other (or both) mesons are off-shell \cite{Janbazi3}.

This review is focus on the method
of three-point QCD sum rules to calculate, the strong
form factors and coupling constants associated with the $B_1B_{s0}K$ and $B_1B_{s1}K$ vertices, for both the $B_{s0}(B_{s1})$ and $K$
states being off-shell.
The three-point correlation function is investigated in two
phenomenological and theoretical sides.
In the physical or phenomenological part, the
representation is in terms of hadronic
degrees of freedom, which is
responsible for the introduction of the form
factors, decay constants, and masses.
In QCD or theoretical part, which consists of two, perturbative
and non-perturbative contributions (In the present work
the calculations contributing the quark-quark and
quark-gluon condensate diagrams are considered as non-perturbative
effects), we
evaluate the correlation function in quark-gluon language and in
terms of QCD degrees of freedom such as, quark
condensate, gluon
condensate, etc, by the help of the Wilson
operator product
expansion(OPE). Equating two sides and
applying the double Borel
transformations, with respect to the momentum
of the initial and final states, to suppress
the contribution of the higher states and
continuum, the strong form factors are estimated.

The outline of the paper is as follows.
In section II, by introducing the sufficient correlation
functions, we obtain QCD sum rules for the
strong coupling constant of the considered
$B_1B_{s0}K$ and $B_1B_{s1}K$
vertices. In obtaining the sum rules for physical
quantities, both light quark-quark
and light quark-gluon condensate diagrams are considered as
non-perturbative contributions. In section III,
the derived sum rules
for the considered strong coupling constants are numerically analyzed with and without $SU_{f}(3)$ symmetry.
We will obtain the numerical values for each
coupling constant when both the $B_{s0}(B_{s1})$ and $K$
states are off-shell. Then taking the average of the
two off-shell cases, we will obtain final numerical
values for each coupling constant.
 In this section, we also compare our results in $SU_{f}(3)$ with the existing
predictions of the other works.
\section{ THE THREE-POINT QCD SUM RULES METHOD}
\begin{figure}[!th]
	\centering
	\includegraphics[width=13cm,height=4cm]{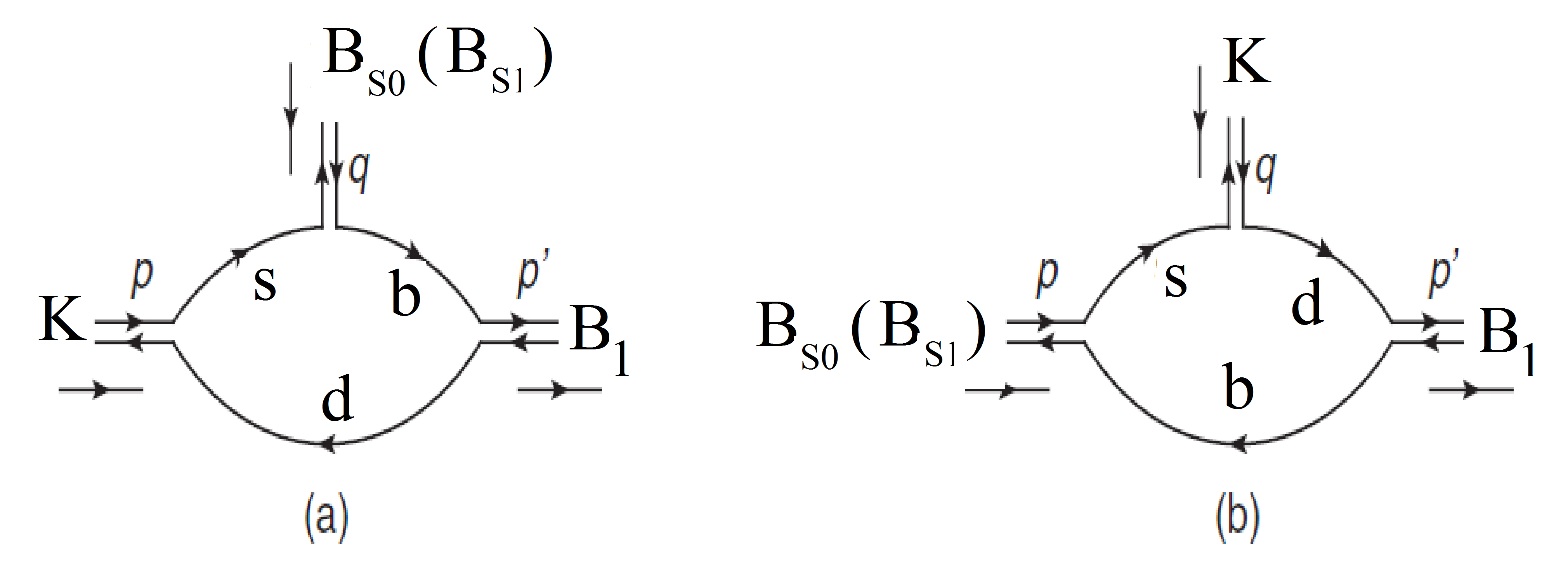}
	\vspace*{-1cm}\caption{perturbative diagrams for off-shell $B_{s0}(B_{s1})$ (left) 	and off-shell $K$ (right).}\label{F1}
\end{figure}

 To evaluate the strong coupling constants, it is necessary to know the effective Lagrangians of the interaction which, for the vertices $ B_{1}B_{s0}K$ and $ B_{1} B_{s1}K$, are\cite{Song12,123}:
\begin{eqnarray}
{\cal L}_{B_1B_{s0}K}&=&-ig_{B_1B_{s0}K} B_1^{\alpha}( \partial_{\alpha}B_{s0}^{-} K^{+}-B_{s0}^{-} \partial_{\alpha} K^{+} )+H.c. , \nonumber \\
{\cal L}_{B_1B_{s1}K}&=&-g_{B_1B_{s1}K} \epsilon^{
\alpha\beta \gamma \sigma}\partial_{\alpha} B_{1\beta}(K^{+} \partial_{\gamma} B^{-}_{s1\sigma}+\partial_{\gamma} B_{{s1}\sigma}^{+} K^{-} ), 
\end{eqnarray}
From these Lagrangians, we can extract elements associated with the  $ B_{1} B_{s0}K$ and $ B_{1} B_{s1}K$  momentum dependent vertices, that can
be written in terms of the form factors:
\begin{eqnarray}\label{eq21}
\langle B_{1}(p', \epsilon') | B_{s0}(q) K(p)\rangle &=&
g_{B_{1} B_{s0}K}(q^2)   \epsilon'.q,\nonumber\\
\langle  B_1(p', \epsilon') |  B_{s1}(q,\epsilon) K(p) \rangle &=&i g_{ B_1B_{s1}K}(q^2) \epsilon^{
\alpha\beta \gamma \sigma}
 \epsilon'_\gamma(p') \epsilon_\sigma(p) p'_\beta q_\alpha,
\end{eqnarray}
where $q=p'-p$, $g_{B_{1}B_{s0}K}(q^2)$, and $g_{B_{1}B_{s1}K}(q^2)$ are the strong form factor $\epsilon$ and $\epsilon'$ are the polarization vector of the $B_{s1}$ and $B_{1}$ mesons. We study the strong coupling constants $ B_{1} B_{s0}K$ and $ B_{1} B_{s1}K$ vertices
when both $K$ and $B_{s0}(B_{s1})$ can be off-shell.
The interpolating currents $j^{K}=\bar d \gamma_5 s$, $j^{B_{s0}}=\bar{s} b$, $j_{\nu}^{B_{s1}}=\bar{s} \gamma_{\nu} \gamma_5 b$ and $j_{\mu}^{B_{1}}=\bar{d} \gamma_{\mu} \gamma_5 b$ are interpolating
currents of $K$, $B_{s0}$, $B_{s1}$, and $B_{1}$ mesons, respectively. We write the three-point correlation function associated with the  $ B_{1}B_{s0}K$ and $ B_{1} B_{s1}K$ vertices. For the off-shell $B_{s0}(B_{s1})$ meson, Fig.\ref{F1} (left), these correlation functions are given
by:

\begin{eqnarray}\label{eq23}
\Pi^{B_{s0}}_{\mu}(p, p')=i^2 \int d^4x d^4y
e^{i(p'x-py)}\langle 0 |\mathcal{T}\left\{{j^{K}}^{\dagger}(x)
{j^{B_{s0}}}^{\dagger}(0) j_{\mu}^{B_{1}}(y)\right\}| 0 \rangle,
\end{eqnarray}
\begin{eqnarray}\label{eq24}
\Pi^{B_{s1}}_{\nu\mu}(p, p')=i^2 \int d^4x d^4y
e^{i(p'x-py)}\langle 0 |\mathcal{T}\left\{{j^{K}}^{\dagger}(x)
{j_{\nu}^{B_{s1}}}^{\dagger}(0) j_{\mu}^{B_{1}}(y)\right\}| 0 \rangle,
\end{eqnarray}
and for the off-shell $K$ meson, Fig.\ref{F1} (right), these quantities
are:
\begin{eqnarray}\label{eq26}
\Pi^{K}_{\mu}(p, p')=i^2 \int d^4x d^4y
e^{i(p'x-py)}\langle 0 |\mathcal{T}\left\{{j^{B_{s0}}}^{\dagger}(x)
{j^{K}}^{\dagger}(0) j_{\mu}^{B_{1}}(y)\right\}| 0 \rangle,
\end{eqnarray}
\begin{eqnarray}\label{eq27}
\Pi^{K}_{\nu\mu}(p, p')=i^2 \int d^4x d^4y
e^{i(p'x-py)}\langle 0 |\mathcal{T}\left\{{j_{\nu}^{B_{s1}}}^{\dagger}(x)
{j^{K}}^{\dagger}(0) j_{\mu}^{B_{1}}(y)\right\}| 0 \rangle,
\end{eqnarray}

Correlation function  in (Eqs. (\ref{eq23} - \ref{eq27}))
in the OPE and in the phenomenological side can be written in terms of several tensor structures. We can write a sum rule to find the coefficients of each structure, leading to as many sum rules as structures. In principle, all the structures should yield the same final results but, the truncation of the OPE changes different structures in different ways. Therefore some structures lead to more stable sum rules. In the $B_1B_{s0}K$ vertex, we have two structures $p'_{\mu}$ and $p_{\mu}$. Two structures give the same
result for $B_1B_{s0}K$. We have chosen the $p'_{\mu}$ structure. In the $B_1B_{s0}K$ vertex, we have only one structure $\epsilon^{ \alpha\beta \mu \nu} p_\alpha p'_\beta$.

With the help of the operator product expansion (OPE) in the Euclidean
region, where $p^2,p'^2\to -\infty$, we calculate the QCD side of
the correlation function (Eqs. (\ref{eq23} - \ref{eq27}))
containing perturbative and non-perturbative parts.
In practice, only the first few condensates contribute significantly, the
most important ones being the 3-dimension, $\langle\bar{d}d\rangle$, and the 5-dimension, $\langle\bar{d}\sigma_{\alpha \beta}
T^{a}G^{a\alpha\beta}d\rangle$, condensates.
For each invariant structure, i, we can write
\begin{eqnarray}\label{eq213}
\Pi^{(theor)}_{i}(p^2, p'^2, q^2) &=& -\frac{1}{4 \pi^2} \int_{(m_d+m_b)^2}^{\infty}
ds'\int_{s_{1(2)}}^{\infty} ds\frac{\rho_{i}(s, s',
	q^2)}{(s-p^2)(s'-p'^2)}\nonumber\\
&+&C^{3}_{i}\langle\bar{d}d\rangle +C_{i}^5\langle\bar{d}\sigma_{\alpha \beta}
T^{a}G^{a\alpha\beta}d\rangle+\cdots,
\end{eqnarray}
where $\rho_i(s, s', q^2)$ is spectral density,
$C_i$ are the Wilson coefficients and $G^{a\alpha\beta}$
is the gluon field strength tensor. We take for the strange quark condensate $\langle\overline{d}d\rangle=- (0.24\pm0.01)^3 ~GeV^3$ \cite{Ioffe} and for the mixed quark-gluon condensate $\langle\bar{d}\sigma_{\alpha \beta}
T^{a}G^{a\alpha\beta}d\rangle=m_0^2\langle\overline{d}d\rangle$ with $m_0^2=(0.8\pm0.2)GeV^2$ \cite{Dosch}.

Furthermore, we make the usual assumption that the contributions of higher resonances are
well approximated by the perturbative expression
\begin{eqnarray}\label{eq214}
-\frac{1}{4 \pi^2} \int_{s'_0}^{\infty}
ds'\int_{s_0}^{\infty} ds\frac{\rho_{i}(s, s',
	q^2)}{(s-p^2)(s'-p'^2)},
\end{eqnarray}
with appropriate continuum thresholds $s_0$, and $s'_0$.

The Cutkosky’s rule allows us to obtain the spectral densities of
the correlation function for the Lorentz
structures appearing in the correlation function. The leading contribution
comes from the perturbative term, shown
in Fig.\ref{F1}.

(i) For the related to the $B_1B_{s0}K $ vertex:
\begin{eqnarray*}
	\rho^{B_{s0}(K)}_{B_1B_{s0}K}&=& 4 N_c I_0\left[A_2\left(m_2m_3-km_1m_2+km_1m_3-m_3^2+\Delta-\frac{u}{2}\right)+km_3^2-m_3m_1-k\frac{\Delta}{2}\right],
\end{eqnarray*}
(ii) For the $\epsilon^{ \alpha\beta \mu \nu} p_\alpha p'_\beta$ structure
related to the $B_1B_{s1}K $ vertex:
\begin{eqnarray*}
	\rho^{B_{s1}(K)}_{B_1B_{s1}K}&=& 4i N_c I_0\left[A_1\left(m_3-km_1\right)+A_2\left(m_2+m_3\right)+m_3\right],
\end{eqnarray*}
The explicit expressions of the coefficients in the spectral densities
entering the sum rules are given as:
\begin{eqnarray*}
	I_0(s,s',q^2) &=& \frac{1}{4\lambda^\frac{1}{2}(s,s',q^2)},\nonumber \\
	\Delta &=& (s+m_3^2-m_1^2),\nonumber \\
	\Delta' &=& (s'+m_3^2-m_2^2),\nonumber \\
	u &=& s+s'-q^2,\nonumber \\
	\lambda(s,s',q^2) &=& s^2+ s'^2+ q^4- 2sq^2- 2s'q^2- 2ss',\nonumber \\
	A_1 &=& \frac{1}{\lambda(s,s',q^2)} \left [2 s' \Delta -\Delta' u\right],\nonumber \\
	A_2 &=& \frac{1}{\lambda(s,s',q^2)} \left [2 s \Delta' -\Delta u\right],
\end{eqnarray*}
Where $k=1$, $m_1=m_s$, $m_2=m_b$, $m_3=m_d$ for $B_{s0}(B_{s1})$ meson off-shell and $k=-1$, $m_1=m_s$, $m_2=m_d$, $m_3=m_b$ for $K$ meson off-shell,
$N_c=3$ represents the color factor.

\begin{figure}[!th]
	\centering
	\includegraphics[width=13cm,height=4cm]{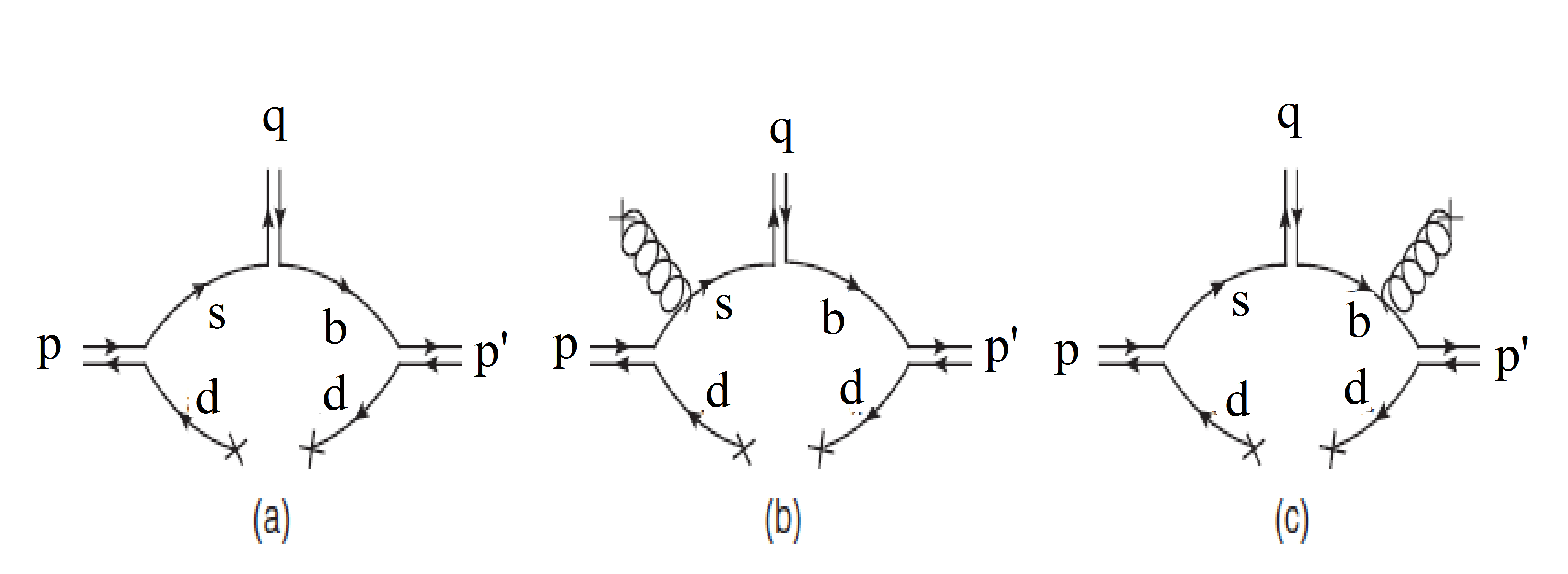}
	\vspace*{-1cm}\caption{Contribution of the
		quark-quark and quark-gluon condensate for the $B_{s0}(B_{s1})$
		off-shell.}\label{F2}
\end{figure}

We proceed to calculate the non-perturbative contributions in the QCD side that
contain the quark-quark and quark-gluon condensate. The quark-quark and quark-gluon condensate
is considered when the light quark is
a spectator \cite{Khodjamirian12};
therefore only three relevant diagrams of dimension 3 and 5 remain from the non-perturbative part contributions when the $B_{s0}(B_{s1})$ meson are off-shell.
These diagrams named quark-quark and quark-gluon condensate are depicted in Fig.\ref{F2}.
For the $K$ off-shell,
there is no quark-quark and quark-gluon condensate contribution.

After some straightforward calculations  and applying the double Borel transformations with respect to
the $p^2(p^2\rightarrow M_1^2)$ and $p'^2(p'^2\rightarrow M_2^2)$ as:
\begin{eqnarray}
{{B}}_{p^2}(M_1^2)(\frac{1}{p^2-m^2_u})^m=\frac{(-1)^m}{\Gamma(m)}
\frac{e^{-\frac{m_u^2}{M_1^2}}}{(M_1^2)^m}, \nonumber \\
{{B}}_{{p^{'}}^2}(M_2^2)(\frac{1}{{p^{'}}^2-m^2_b})^n=\frac{(-1)^n}{\Gamma(n)}
\frac{e^{-\frac{m_b^2}{M_2^2}}}{(M_2^2)^n},
\end{eqnarray}
where $M_1^2$ and $M_2^2$ are the Borel parameters,
the contributions of the quark-quark and quark-gluon condensate 
for the $B_{s0}(B_{s1})$ meson off-shell case, are given by:
\begin{eqnarray}\label{eq218}
\Pi_{(non-per)}^{B_{s0}(B_{s1})}&=&\langle\overline{d}d\rangle~\frac{C^{B_{s0}(B_{s1})}}{M_1^4M_2^4},
\end{eqnarray}
The explicit expressions for $C_{B_1B_{s0}K(B_1B_{s1}K)}^{B_{s0}(B_{s1})}$ associated
with the $B_1 B_{s0}K$ and $B_1 B_{s1}K$
vertices are given in the appendix.

The gluon-gluon condensate is considered when the heavy quark is a spectator \cite{Likhoded}, and the $B_{s0}(B_{s1})$ mesons are off-shell,
and there is no gluon-gluon condensate contribution.
Our numerical analysis shows that the contribution of the non-perturbative part containing the quark-quark and quark-gluon diagrams is about $13\%$  and the gluon-gluon contribution is about $3\%$  of the total, and the main contribution comes from the perturbative part of the strong form factors, and we can ignore gluon-gluon contribution in our calculation\cite{Guo123, Janbazi3}.

The phenomenological side of the vertex function is obtained
by considering the contribution of three complete sets of
intermediate states with the same quantum number that should
be inserted in Eqs. (\ref{eq23} - \ref{eq27}).
We use the standard definitions for the decay constants
$f_M$ ($ f_{K} $, $f_{B_{s0}}$, $f_{B_{s1}}$,  and $f_{B_{1}}$) and are given by:
\begin{eqnarray}\label{eq29}
\langle 0 | j^{K} | K(p) \rangle &=& \frac{m_{K}^2 f_{K}}{m_s+m_d}, \nonumber \\
\langle 0 | j^{B_{s0}} | B_{s0}(p) \rangle &=& m_{B_{s0}} f_{B_{s0}},\nonumber \\
\langle 0 | j_{\nu}^{B_{s1}} | {B_{s1}}(p, \epsilon) \rangle &=&
m_{B_{s1}} f_{B_{s1}} \epsilon_{\nu}(p),\nonumber\\
\langle 0 | j_{\mu}^{B_{1}} | {B_{1}}(p', \epsilon') \rangle &=&
m_{B_{1}} f_{B_{1}} \epsilon'_{\mu}(p'),
\end{eqnarray}
The phenomenological part for the $p'_{\mu} $ structure
 related to the $B_1B_{s0}K $ vertex, when $B_{s0}(K)$ is off-shell  meson is:
 \begin{eqnarray}\label{eq211}
\Pi^{B_{s0}(K)}_{\mu}&=&-g_{B_{1} B_{s0}K}^{B_{s0}(K)}(q^2)
\frac{ m_{K}^2m_{B_{s0}}m_{B_1} f_{K} f_{B_{s0}} f_{B_{1}}(m_{B_{1}}^2+
m_{K(B_{s0})}^2-q^2)}{2(q^2-m_{B_{s0}(K)}^2)(p^2-m_{K(B_{s0})}^2)
(p'^2-m_{B_{1}}^2)(m_s+m_d)} p'_{\mu}+h.r, \nonumber \\
\end{eqnarray}
The phenomenological part for the $\epsilon^{ \alpha\beta \mu \nu} p_\alpha p'_\beta$ structure
 related to the $B_1B_{s1}K $ vertex, when $B_{s1}(K)$ is off-shell  meson is:
\begin{eqnarray}\label{eq212}
\Pi^{B_{s1}(K)}_{\mu\nu}&=&-i g_{B_1 B_{s1} K}
^{B_{s1}(K)}(q^2)\frac{m_{K}^2 m_{B_1} m_{B_{s1}}   f_{K} f_{B_1}f_{B_{s1}} }{(q^2-m_{B_{s1}(K)}^2)
(p^2-m_{K(B_{s1})}^2)(p'^2-m_{B_1}^2)(m_s+m_d)}\epsilon^{ \alpha\beta \mu \nu} p_\alpha p'_{\beta}+h.r,\nonumber\\
\end{eqnarray}
In the Eqs.(\ref{eq211} - \ref{eq212}), h.r. represents the
contributions of the higher states and continuum.

The QCD sum rules for the strong form factors are obtained
after performing the Borel transformation with respect
to the variables $p^2 (B_{p^2}(M_1^2))$ and $p'^2(B_{p'}^2(M_2^2))$
on the physical (phenomenological) and QCD parts and equating  these
two representations of the correlations, we obtain the corresponding equations for
the strong form factors as follows.

$\bullet$ For the $g_{B_1B_{s0}K}(Q^2)$ form factors:
\begin{eqnarray}\label{eq220}
g^{B_{s0}}_{B_1 B_{s0}K}(Q^2)&=& \frac{2(Q^2+m^2_{B_{s0}})(m_s+m_d)}{m_{K}^2 m_{B_{s0}}m_{B_1}
f_{K}f_{B_{s0}}f_{B_1}(m_{B_1}^2+m_{K}^2+Q^2)} ~e^{\frac{m_{K}^2}{M_1^2}}
e^{\frac{m_{B_1}^2}{M_1^2}}
 \left\{-\frac{1}{4\pi^2}\int^{s'_0}_{(m_b
+m_d)^2} ds'\right. \nonumber \\ &&
\left. \times \int^{s_0}_{s_{1}} ds \rho^{B_{s0}}(s,s',Q^2)
e^{-\frac{s}{M_1^2}} e^{-\frac{s'}{M_2^2}}+\langle d\bar d \rangle\frac{C_{B_1B_{s0}K}^{B_{s0}}}{M_1^2M_2^2}
\right\},
\end{eqnarray}
\begin{eqnarray}\label{eq223}
g^{K}_{B_1B_{s0}K}(Q^2)&=& \frac{2(Q^2+m^2_{K})(m_s+m_d)}{m_{K}^2 m_{B_{s0}}m_{B_1}
f_{K}f_{B_{s0}}f_{B_1}(m_{B_1}^2+m_{B_{s0}}^2+Q^2)} ~e^{\frac{m_{B_{s0}}^2}{M_1^2}}
e^{\frac{m_{B_1}^2}{M_2^2}}
 \left\{-\frac{1}{4\pi^2}\int^{s'_0}_{(m_b
+m_d)^2} ds'\right. \nonumber \\ &&
\left. \times \int^{s_0}_{s_{2}} ds \rho^{K}(s,s',Q^2)
e^{-\frac{s}{M_1^2}} e^{-\frac{s'}{M_2^2}}\right\},
\end{eqnarray}

$\bullet$ For the $g_{B_1B_{s1}K}(Q^2)$ form factors:
\begin{eqnarray}\label{eq221}
g^{B_{s1}}_{B_1B_{s1}K}(Q^2)&=& -i \frac{(Q^2+m^2_{B_{s1}})(m_s+m_d)}{m_{K}^2 m_{B_1} m_{B_{s1}}
f_{K}f_{B_1} f_{B_{s1}}} ~e^{\frac{m_{K}^2}{M_1^2}}
e^{\frac{m_{B_1}^2}{M_2^2}}
 \left\{-\frac{1}{4\pi^2}\int^{s'_0}_{(m_b
+m_d)^2} ds'\right. \nonumber \\ &&
\left. \times \int^{s_0}_{s_{1}} ds \rho^{B_{s1}}(s,s',Q^2)
e^{-\frac{s}{M_1^2}} e^{-\frac{s'}{M_2^2}}+\langle d\bar d \rangle\frac{C_{B_1B_{s1}K}^{B_{s1}}}{M_1^2M_2^2}
\right\},
\end{eqnarray}

\begin{eqnarray}\label{eq224}
g^{K}_{B_1B_{s1}K}(Q^2)&=& -i\frac{(Q^2+m^2_{K})(m_s+m_d)}{m_{K}^2 m_{B_1}m_{B_{s1}}
f_{\pi}f_{B_1}f_{B_{s1}}} ~e^{\frac{m_{B_{s1}}^2}{M_1^2}}
e^{\frac{m_{B_1}^2}{M_2^2}}
 \left\{-\frac{1}{4\pi^2}\int^{s'_0}_{(m_b
+m_d)^2} ds'\right. \nonumber \\ &&
\left. \times \int^{s_0}_{s_{2}} ds \rho^{K}(s,s',Q^2)
e^{-\frac{s}{M_1^2}} e^{-\frac{s'}{M_2^2}}\right\},
\end{eqnarray}
where $Q^2=-q^2$, $ s_0 $ and $s'_0  $ are the continuum
thresholds, and $s_1$ and $s_2$ are the lower limits of the integrals over $s$ as:
\begin{eqnarray}\label{eq225}
s_{1(2)}=\frac{(m_{d(b)}^{2}+q^2-m_{s}^{2}-s')
(m_{s}^{2}s'-q^2m_{d(b)}^{2})}{(m_{s}^{2}-q^2)(m_{d(b)}^{2}-s')}~.
\end{eqnarray}
\section{NUMERICAL ANALYSIS}
In this section, numerical analysis for the expressions of the strong coupling constant is presented. The values of masses for quarks and mesons are given in Table \ref{mass}. The leptonic decay constants used in these calculations are shown in Table \ref{T1}.                       

\begin{table}[th]
	\caption{The
		values of quark and meson masses in $\rm GeV$ \cite{PDG2012}.}\label{mass}
	\begin{ruledtabular}
		\begin{tabular}{cccccc}
			$m_s$& $m_b$&$m_{K}$&$m_{B_{s0}}$& $m_{B_{s1}}$&$m_{B_1}$ \\
			\hline
			$0.14\pm0.01$& $4.67\pm0.1$& $0.493$ & $5.70$&$5.72$& $5.72$
		\end{tabular}
	\end{ruledtabular}
\end{table}

\begin{table}[th]
	\caption{The leptonic decay constants in $\rm MeV$.}\label{T1}
	\begin{ruledtabular}
		\begin{tabular}{cccc}
			$f_{K}$\cite{fmeson}& $f_{B_1}$\cite{Bazavov}&  $f_{B_0}$\cite{THuang}& $f_{B_{s1}}$\cite{Wang}\\
			\hline
			$156.1\pm8$& $196.9 \pm 8.9$& $280\pm31$& $240\pm20$
		\end{tabular}
	\end{ruledtabular}
\end{table}

There are four auxiliary parameters containing the Borel mass parameters M$_1^2$ and M$_2^2$, and continuum thresholds $s^{K}_{0}$, $s_{0}^{B_1}$, $s_{0}^{B_{s0}}$ and $s_{0}^{B_{s1}}$ in Eqs.(\ref{eq220}-\ref{eq224}). The coupling constants  and strong form factors 
as  physical quantities should be independent of the auxiliary parameters. Howeve, the continuum thresholds are not arbitrary entirely; these are related to the energy of the first excited state. The values of the continuum thresholds are taken to be $s^{K}_{0}=(m_{K}+\delta)^2$, $s_{0}^{B_{s0}}=(m_{B_{s0}}+\delta')^2$, $s_{0}^{B_{s1}}=(m_{B_{s1}}+\delta')^2$ and $s_{0}^{B_1}=(m_{B_1}+\delta')^2$. We use $0.50 ~\rm GeV^2\leq
\delta \leq 0.90~\rm \rm GeV^2$ and $0.30 ~\rm GeV^2\leq \delta'
\leq0.70~\rm \rm GeV^2$ \cite{Janbazi3}.

\begin{figure}[!th]
	\centering
	\includegraphics[width=8cm,height=8cm]{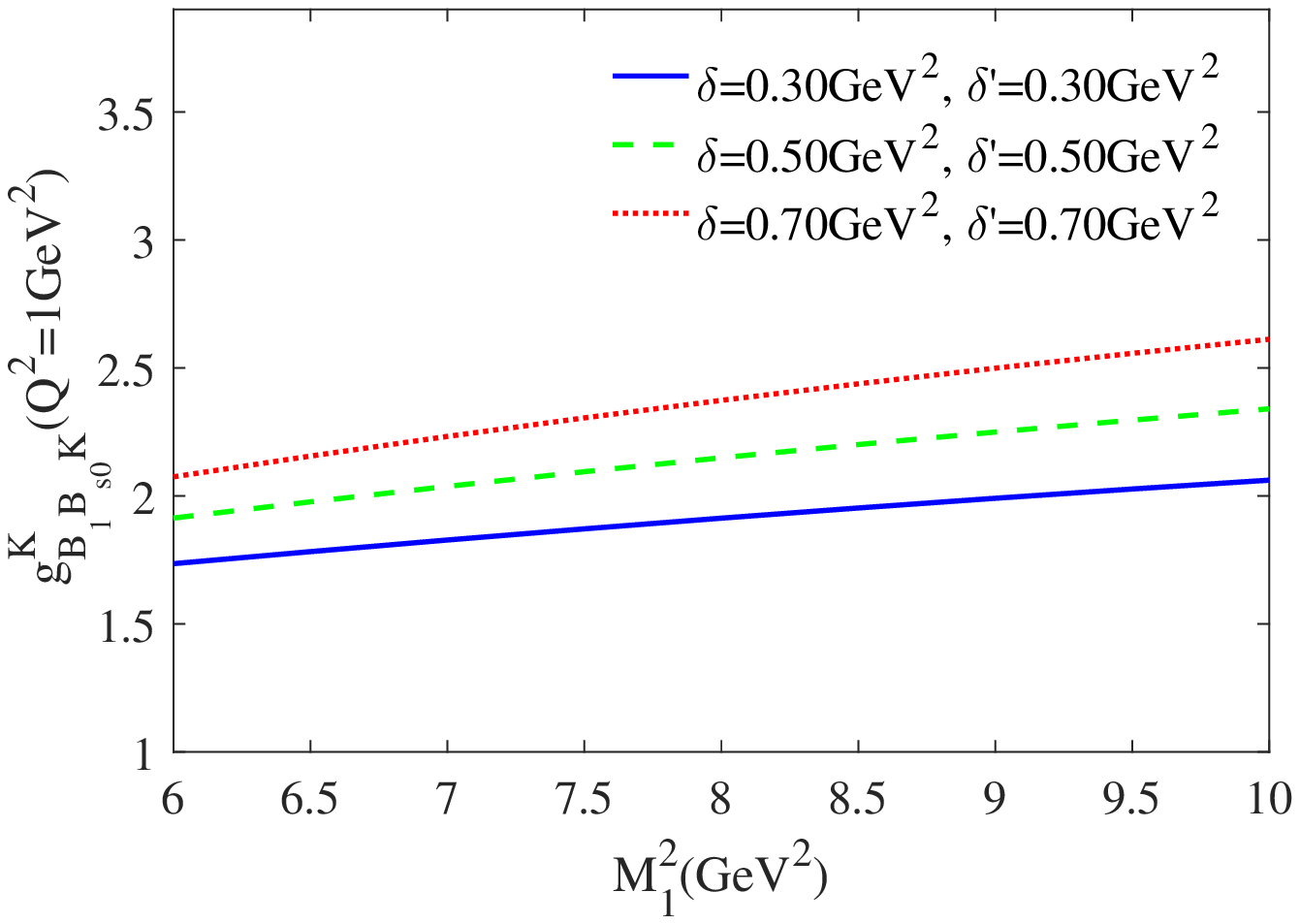}\includegraphics[width=8cm,height=8cm]{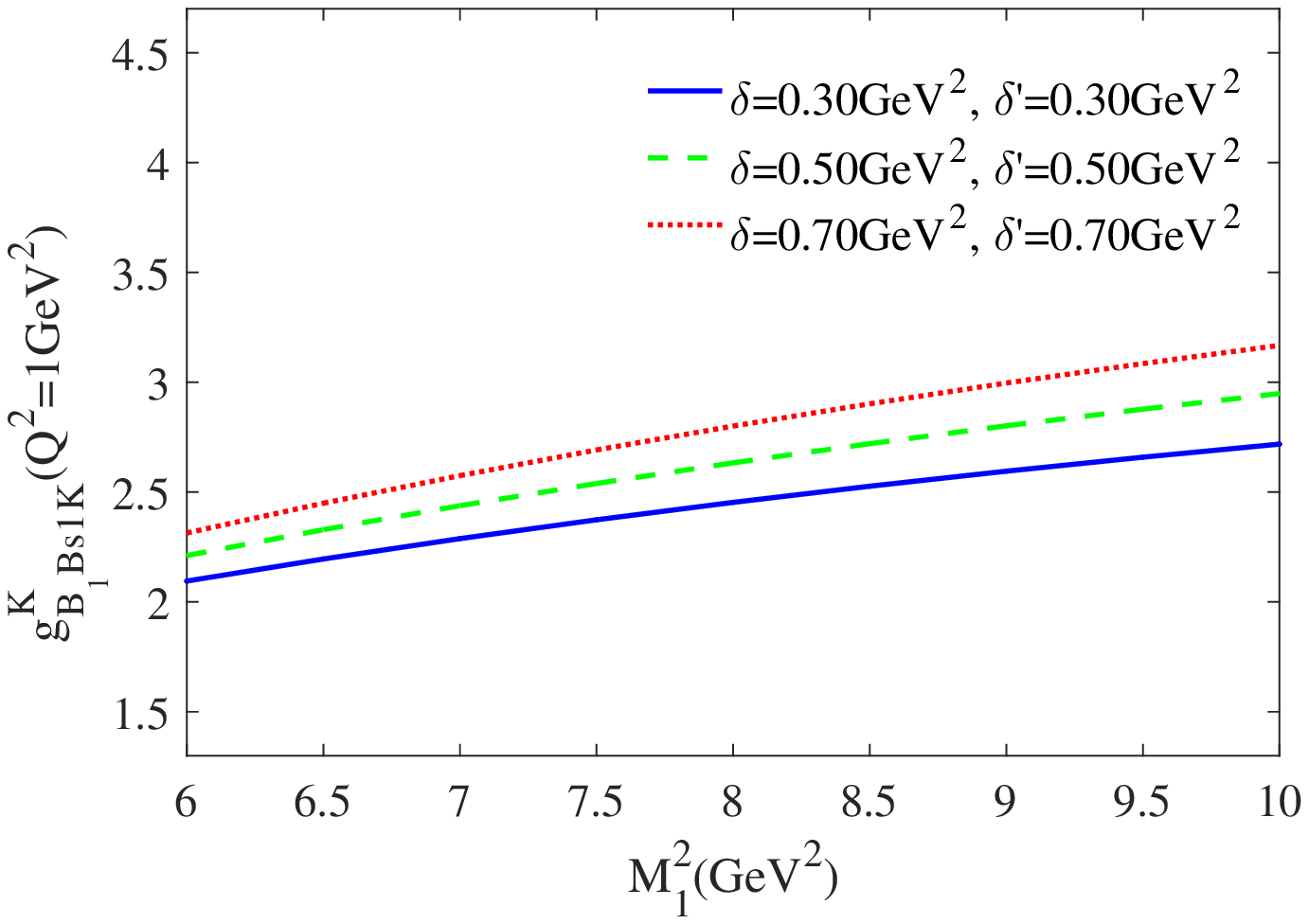}
	\caption{The strong form factors $g^{K}_{B_{1} B_{s0}K}$  (left) and $g^{K}_{ B_{1}  B_{s1}K}$ (right) as functions of the Borel mass parameter $M_1^2$.}
	\label{F301}
\end{figure}

\begin{figure}[!th]
	\centering
	\includegraphics[width=8cm,height=8cm]{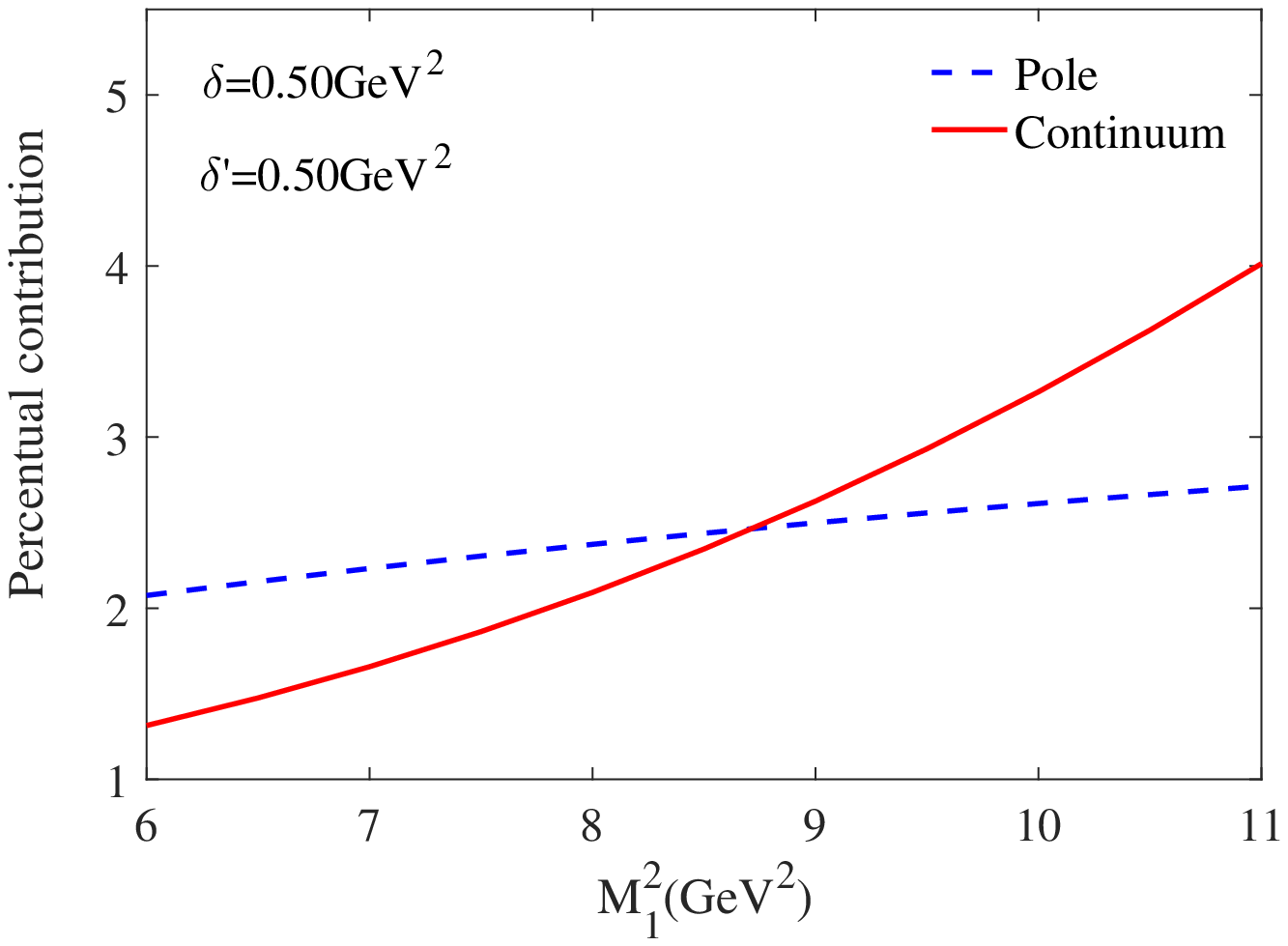}\includegraphics[width=8cm,height=8cm]{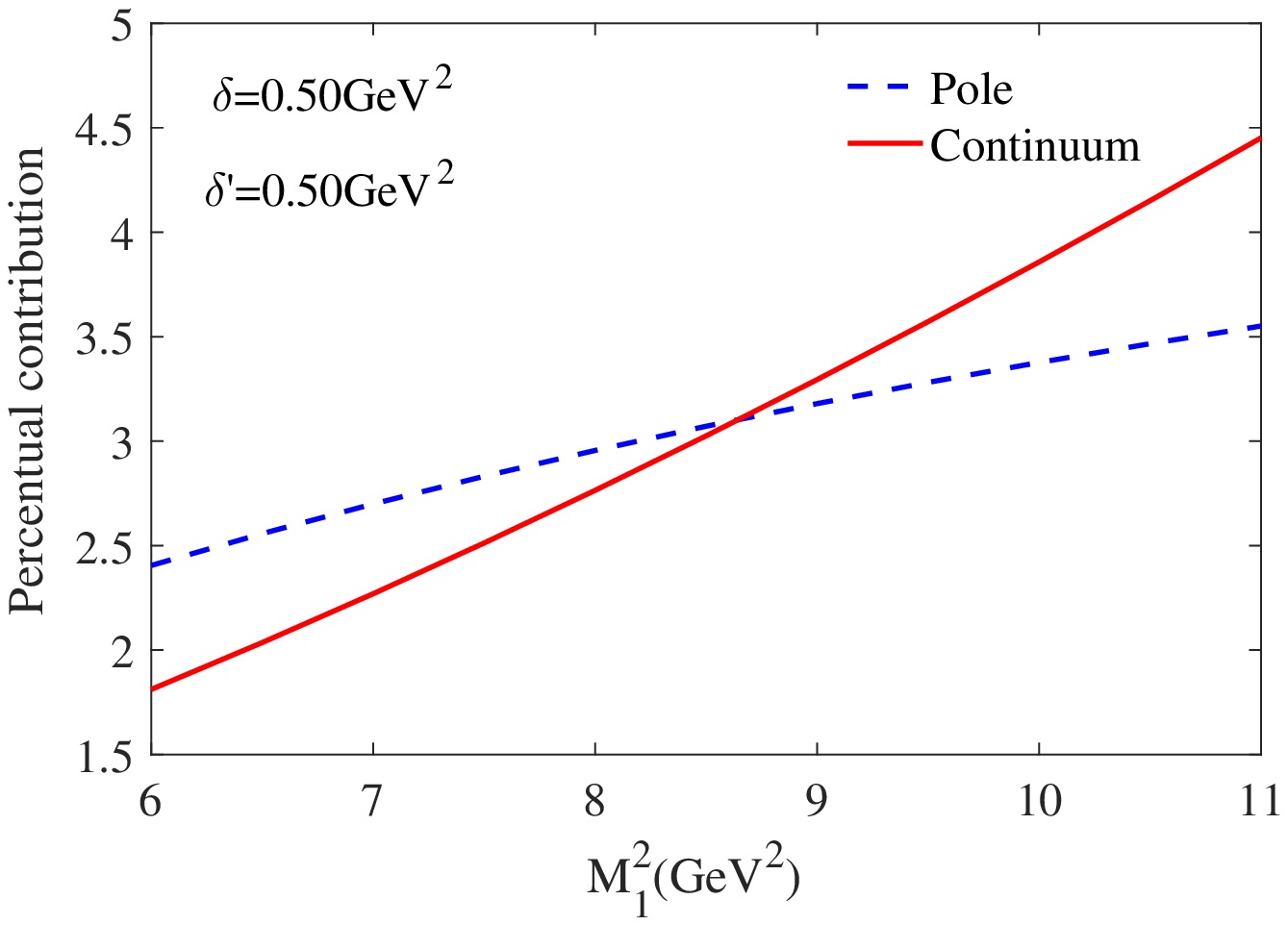}
	\caption{Pole and continuum contributions for the strong form factors $g^{K}_{B_{1} B_{s0}K}$  (left) and $g^{K}_{ B_{1}  B_{s1}K}$ (right)  as functions of the Borel mass parameter $M_1^2$.}
	\label{F302}
\end{figure}

Our results should be almost insensitive to the Borel parameters intervals. 
On the other hand, the intervals of the Borel mass parameters must suppress the higher states, continuum, and contributions of the highest-order operators. In other words, the sum rule for the strong form factors must converge and the stability of our results \cite{Guo123, Bracco123}. This interval is called the “Borel window.”
In this work, the following relations between the Borel masses $M_1^2$ and $ M_2^2 $ is  $ \frac{M_1^2}{M_2^2}=\frac{m_{K}^2}{m_{B_1}^2-m_{b}^2} $ when $B_{s0}(B_{s1})$ meson is off-shell and $M_1^2=M_2^2 $ when $K$  meson is off-shell. We have illustrated the form factors of $ B_{1} B_{s0}K$ and $ B_{1}  B_{s1}K$  vertices for $K$ off-shell respect to the Borel parameter $M_1^2$ for three values of the continuum thresholds  $s^{K}_{0}$ and $s_{0}^{B_1}$ are shown in Figure \ref{F301}. In Figure \ref{F302}, we also show the pole-continuum analysis for the strong form factors $g^{K}_{B_{1}  B_{s0}K}$ and $g^{K}_{ B_{1} B_{s1}K}$. As it can be seen, for $M_1^2 <9 ~{\rm GeV^2} $  the sum rule is dominated by the pole contribution for the strong form factors $g^{K}_{B_{1} B_{s0}K}$ and $g^{K}_{ B_{1} B_{s1}K}$. Thus, we choose a Borel window where the pole contribution is between $20\%$ and $80\%$ of the QCDSR total contribution what we choose the
interval $6 ~{\rm GeV^2} < M_1^2 < 11 ~{\rm GeV^2} $  for the strong form factors $g^{K}_{B_{1}  B_{s0}K}$ and $g^{K}_{ B_{1}B_{s1}K}$. According to the same analysis with $K$ off-shell, we choose the Borel window $8 ~{\rm GeV^2} < M_1^2 < 12 ~{\rm GeV^2} $  ($Q^2 = 3.0 GeV^2$) for the strong form factors $g^{B_{s0}}_{B_{1}  B_{s0}K}$ and $g^{B_{s1}}_{ B_{1}  B_{s1}K}$. 

\begin{figure}[!th]
	\includegraphics[width=8cm,height=8cm]{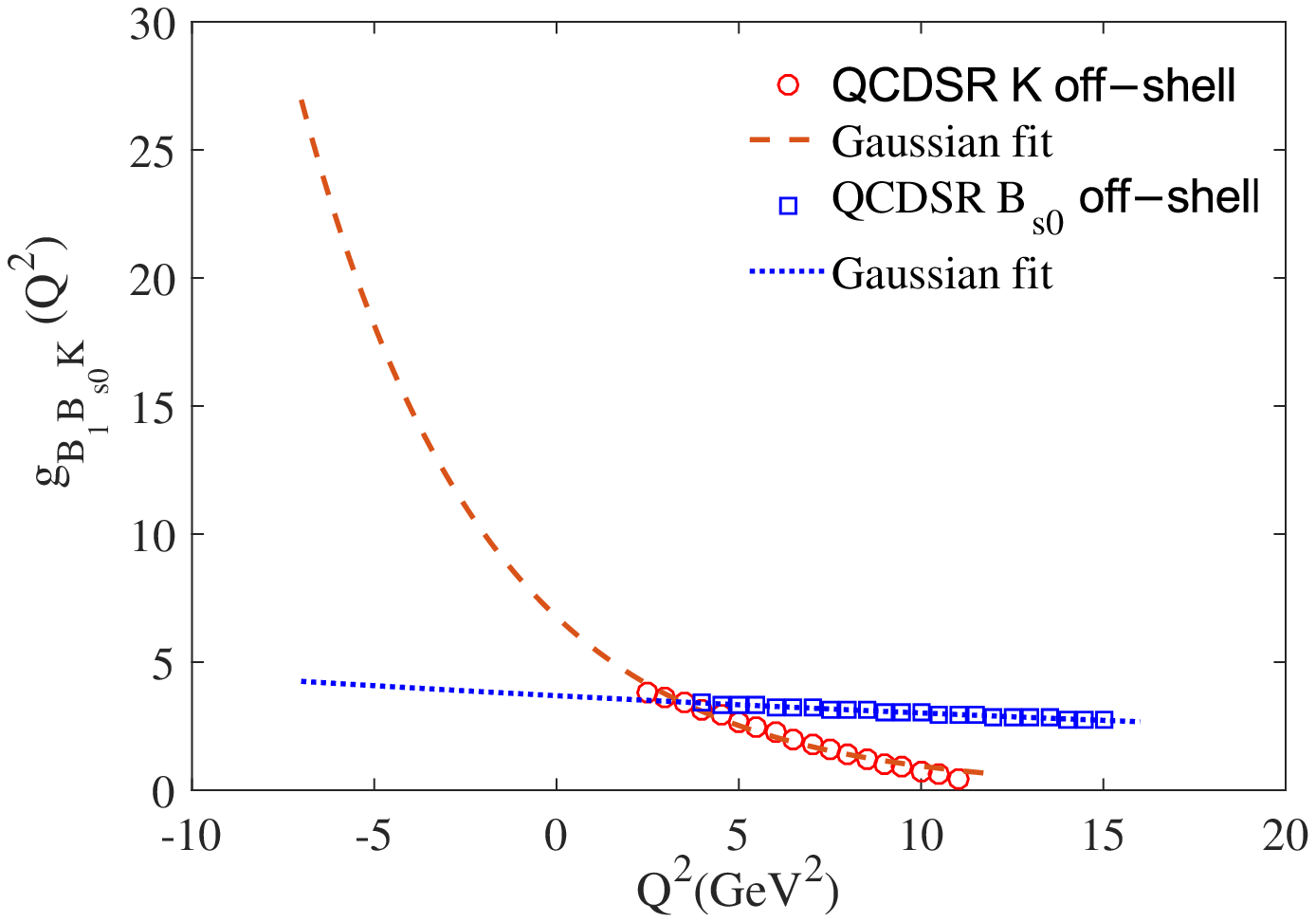}\includegraphics[width=8cm,height=8cm]{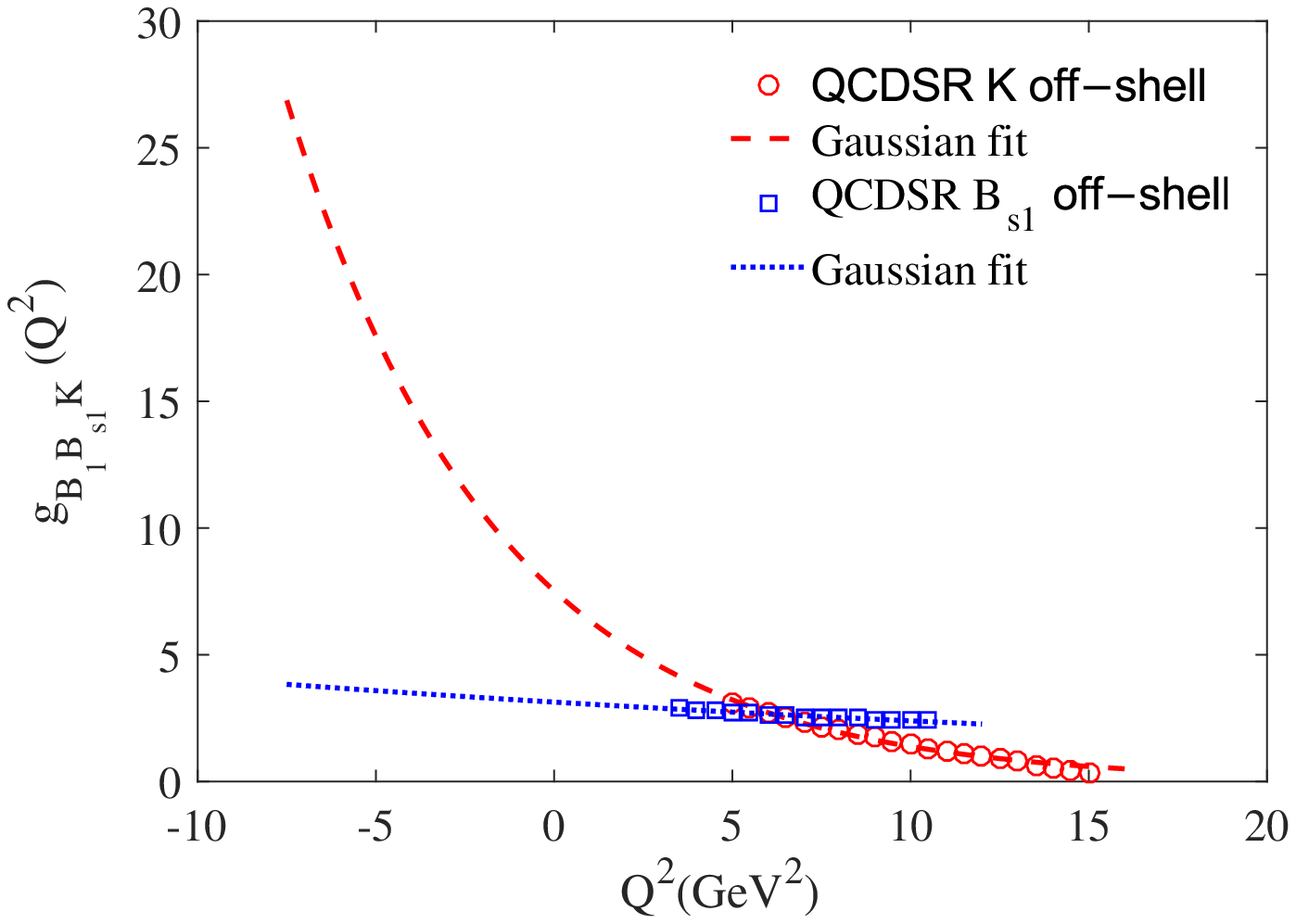}
	\caption{The strong form factors $g_{B_1B_{s0}K}$ and $g_{B_1B_{s1}K}$ on $Q^2$ (The boxes and circles the results of the numerical evaluation via the 3PSR for the form factors).}
	\label{F32}
\end{figure}

We have chosen the Borel mass to be $ M_1^2= 7~GeV^2 $ and $ M_1^2= 9~GeV^2 $ for off-shell $K$  and $B_{s0}(B_{s1})$, respectively.
Having determined $M_1^2 $, we calculated the $Q^2$
dependence of the form factors. We present the results in
Fig.\ref{F32} for the  $g_{B_1B_{s0}K}$ and $g_{B_1B_{s1}K}$ vertices.
In these figures, the small circles and boxes correspond to the form factors in the interval where the sum rule is valid. As it is seen, the form factors and their fit functions coincide together, well.

We discuss a difficulty inherent to the calculation of coupling
constants with QCDSR. The solution of Eqs. (\ref{eq220}-\ref{eq224}) are numerical and restricted to a singularity-free region in
the $Q^2 $ axis, usually located in the space-like region. Therefore, in order to reach the pole position, $Q^2 = -m_m^2$, we must fit the solution by finding a function $g(Q^2)$, which is then extrapolated to the pole yielding the coupling constant.

The uncertainties associated with the extrapolation procedure, for each vertex is minimized by performing the calculation twice, first  putting one meson and then another meson off-shell, to obtain two form factors $g^{B_{s0}(B_{s1})}$ and   $g^{K}$ and equating these two functions
at the respective poles.

we find that the sum rule predictions of the form factors in Eqs. (\ref{eq220}-\ref{eq224}) are well fitted to the following function:

\begin{eqnarray}\label{eq33}
	g(Q^2)=A~e^{-Q^2/B}.
\end{eqnarray}

The values of the parameters $A$ and $B$ are given in Table
\ref{T311}.

\begin{table}
	\caption{Appeared parameters  in the fit functions of the
		B$_1$B$_{s0}$K and B$_1$B$_{s1}$K, vertices for various 	$(\delta, \delta')$, where  $(\delta_1,\delta'_1)=[0.30(50),0.30(0.30)],
		~(\delta_2,\delta'_2)=[0.50(70),0.50(0.50)]$ and
		$(\delta_3,\delta'_3)=[0.70(90),0.70(0.70)] ~\rm GeV^2$ for $K$ [$B_{s0}(B_{s1})$] off-shell.}\label{T311}
	\begin{ruledtabular}
		\begin{tabular}{ccccccc}
			\hline
			$\mbox{Form factor}$&$A(\delta_1,\delta'_1)$&$B(\delta_1,\delta'_1)$&$A(\delta_2,\delta'_2)$&$B(\delta_2,\delta'_2)$&$A(\delta_3,\delta'_3)$&$B(\delta_3,\delta'_3)$\\
			\hline
			$g^{K}_{B_1B_{s0}K}(Q^2)$&6.57&5.55&6.76&5.06&7.15&4.56 \\
			$g^{B_{s0}}_{B_1B_{s0}K}(Q^2)$&3.58&53.14&3.68&49.53&3.89&48.56 \\
			$g^{K}_{B_1B_{s1}K}(Q^2)$&7.06&6.46&7.52&5.89&7.74&4.87 \\
			$g^{B_{s1}}_{B_1B_{s1}K}(Q^2)$&2.96&37.71&3.16&37.07&3.26&35.88 \\
		\end{tabular}
	\end{ruledtabular}
\end{table}

We define the coupling constant as the value of the strong
coupling form factor at $Q^2 = -m_m^2$ in the Eq. (\ref{eq33}), where $m_m$ is the mass of the off-shell meson. Considering the uncertainties result with the continuum threshold and uncertainties
in the values of the other input parameters, we obtain the average values of the strong coupling
constants shown in Table \ref{T32}.

We can see that for the two cases considered here, the off-
shell  $K$ and $B_{s0}(B_{s1})$ meson,  give compatible results for the
coupling constant.

\begin{table}[th]
\caption{The strong coupling constants $g_{B_1B_{s0}K}$ and
$g_{B_1B_{s1}K}$ . }\label{T32}
\begin{ruledtabular}
\begin{tabular}{cccc}
$\mbox{Coupling constant}$&$\mbox{off-shell $B_{s0}(B_{s1})$}$&$\mbox{off-shell K}$&$\mbox{Average}$ \\
\hline
$g_{B_1B_{s0}K}$&$7.13\pm0.52$&$7.10\pm0.46$&$7.12\pm0.52$\\
$g_{B_1B_{s1}K}(GeV^{-1})$&$7.65\pm0.51$&$7.83\pm0.34$&$7.74\pm 0.51$\\
\end{tabular}
\end{ruledtabular}
\end{table}
In order to investigate the strong coupling constant value via the
$SU_{f}(3)$ symmetry, the mass of the $s$ quark is ignored in all
equations. In view of the $SU_{f}(3)$ symmetry, the values of the
parameters $A$ and $B$ for the $g_{B_1B_{s0}K}$ and
$g_{B_1B_{s1}K}$ vertices in
$(\delta,\delta')=[0.50(70),0.50(0.50)]~\rm GeV^2$ are given in Table \ref{T34}.

\begin{table}[th]
	\caption{Parameters appearing in the fit functions for the
		$g_{B_1B_{s0}K}$ and $g_{B_1B_{s1}K}$ form
		factors in $SU_{f}(3)$ symmetry with $(\delta,\delta')=[0.50(70),0.50(0.50)]~\rm GeV^2$.}\label{T34}
	\begin{ruledtabular}
		\begin{tabular}{ccccccc}
			$\mbox{Form factor}$&$A$&$B$&$\mbox{Form factor}$&$A$&$B$\\
			\hline 
			$g^{K}_{B_1 B_{s0}K}(Q^2)$&4.69&5.03&$g^{B_{s0}}_{B_1B_{s0}K}(Q^2)$&2.17&42.31&\\
			$g^{K}_{B_1  B_{s1}K}(Q^2)$&3.01&5.84&$g^{B_{s1}}_{B_1B_{s1}K}(Q^2)$&1.29&34.84&\\
		\end{tabular}
	\end{ruledtabular}
\end{table}

In addition, considering the $SU_{f}(3)$ symmetry, we obtain the values of the coupling constants of the vertices $B_1B_{s0}K$ and $B_1B_{s1}K$, as shown in Table \ref{T39}. 

\begin{table}[th]
	\caption{The strong coupling constants $g_{B_1B_{s0}K}$ and
		$g_{B_1B_{s1}K}$ in $SU_{f}(3)$ symmetry. }\label{T39}
	\begin{ruledtabular}
		\begin{tabular}{cccc}
			$\mbox{Coupling constant}$&$\mbox{off-shell $B_{s0}(B_{s1})$}$&$\mbox{off-shell K}$&$\mbox{Average}$ \\
			\hline
			$g_{B_1B_{s0}K}$&$4.71\pm0.42$&$4.92\pm0.36$&$4.82\pm0.42$\\
			$g_{B_1B_{s1}K}(GeV^{-1})$&$3.30\pm0.46$&$3.14\pm0.37$&$3.22\pm0.45$\\
		\end{tabular}
	\end{ruledtabular}
\end{table}

It is possible to compare the coupling constant values of $g_{B_1B_{s0}K}$ and $g_{B_1B_{s1}K}$  with $g_{B_0B_1 \pi}$ and $g_{B_1B_1 \pi}$, respectively, in the $SU_{f}(3)$ symmetry consideration. Table \ref{T310} shows a comparison between our results with the findings of others, previously calculated. From this Table, we see that our result of the coupling constants is in a fair agreement with the calculations in refs.\cite{Janbazi1, Zhu123}.

\begin{table}[th]
	\caption{Comparison of our results for strong coupling constants $g_{B_1B_{s0} K }$ and 	$g_{ B_1B_{s1} K}$ in $SU_{f}(3)$ symmetry with the other published results.}\label{T310}
	\begin{ruledtabular}
		\begin{tabular}{cccc}
			$\mbox{Coupling constant}$&$\mbox{Our result}$&$\mbox{3PSR \cite{Janbazi1}}$&$\mbox{LCSR \cite{Zhu123} }$ \\
			\hline
			$g_{B_1B_{s0} K }$&$4.82\pm0.42$&$5.29\pm1.40$&$4.73\pm1.14$\\
			$g_{B_1B_{s1} K }(GeV^{-1})$&$3.22\pm0.4$&$3.57\pm0.53$&$2.60\pm0.60$\\
		\end{tabular}
	\end{ruledtabular}
\end{table}

In summary, in this article, we analyzed the vertices $B_1B_{s0} K $ and $B_1B_{s1} K $ within the framework of the three-point QCD sum rules approach in a unified way. The strong coupling constants could give useful information about strong interactions of the strange $B_{s0}(B_{s1})$ and strange K mesons and also give useful information about the structure of the axial vector and scalar $B_{s0}(B_{s1})$ mesons.

\appendix
\begin{center}
{ \textbf{Appendix: NON-PERTURBATIVE CONTRIBUTIONS }}
\end{center}
\setcounter{equation}{0} \renewcommand{\theequation}

In this appendix,  the explicit expressions of the coefficients of
the quark-quark and quark-gluon  condensate of the strong form
factors for the vertices  $B_1B_{s0} K $ and $B_1B_{s1} K $
with applying the double Borel transformations
are given.

\begin{eqnarray*}
C_{B_1B_{s0} K }^{B_{s0}} &=&(\frac{M_1^2m_0^2m_b}{4}-\frac{M_1^2m_b^2m_d}{2}-\frac{M_1^2m_bm_d^2}{2}-\frac{3m_0^2M_2^2m_s}{4}-M_1^2M_2^2m_s+\frac{m_0^2m_b^2m_s}{4}\\&&-\frac{M_1^2m_bm_dm_s}{2}+\frac{M_1^2m_d^2m_s}{2}-\frac{m_b^2m_d^2m_s}{2}-\frac{M_1^2m_dm_s^2}{2}-\frac{M_2^2m_dm_s^2}{2}+\frac{M_2^2m_d^2mu}{2}\\&&+\frac{m_0^2m_s^3}{4}-\frac{m_d^2m_s^3}{2}+\frac{M_1^2m_dq^2}{2}-\frac{m_0^2m_sq^2}{4}+\frac{m_d^2m_sq^2}{2})\times
e^{-\frac{m_s^2}{M_1^2}}~e^{-\frac{m_b^2}{M_2^2}},
\end{eqnarray*}
\begin{eqnarray*}
C_{B_1B_{s1} K }^{B_{s1}} &=&i(\frac{7m_0^2M_1^2}{12}+\frac{3m_0^2M_2^2}{4}+M_1^2M_2^2-
\frac{m_0^2m_b^2}{2}-\frac{M_1^2m_bm_d}{2}-M_1^2m_d^2-M_2^2m_d^2\\ &&+\frac{M_2^2m_dm_s}{2}-\frac{m_0^2m_s^2}{2}+\frac{m_0^2q^2}{2}-m_d^2q^2+m_b^2m_d^2)\times
e^{-\frac{m_s^2}{M_1^2}}~e^{-\frac{m_b^2}{M_2^2}},
\end{eqnarray*}

\end{document}